\documentclass[pra,twocolumn,aps,10pt,showpacs]{revtex4-1}

\usepackage{graphicx}
\usepackage{verbatim}
\usepackage{comment}
\usepackage{amsmath}
\usepackage{amssymb}
\usepackage{stmaryrd}
\usepackage{color}

\begin{document}

\title{Effect of particle statistics in strongly correlated two-dimensional 
Hubbard models}

\author{Ehsan Khatami and Marcos Rigol} 
\affiliation{Department of Physics, Georgetown University, Washington
District of Columbia, 20057, USA}
\affiliation{Physics Department, The Pennsylvania State University, 104 Davey Laboratory, 
University Park, Pennsylvania 16802, USA}

\begin{abstract}
We study the onset of particle statistics effects as the temperature is lowered
in strongly correlated two-dimensional Hubbard models. We utilize numerical linked-cluster 
expansions and focus on the properties of interacting lattice fermions and two-component 
hard-core bosons. In the weak-coupling regime, where the ground state of the bosonic system 
is a superfluid, the thermodynamic properties of the two systems at half filling
exhibit very large differences even at high temperatures. In the strong-coupling regime, 
where the low-temperature behavior is governed by a Mott insulator for either 
particle statistics, the agreement between the thermodynamic properties of both 
systems extends to regions where the antiferromagnetic (iso)spin correlations are
exponentially large. We analyze how particle statistics affects adiabatic 
cooling in those systems.
\end{abstract}


\maketitle

\section{introduction}

The Fermi-Hubbard model has been the {\em de facto} playground 
for exploring the properties of high-temperature superconductors
for more than two decades~\cite{p_anderson_87}. Yet, still no analytical 
solution exists in more than one dimension, and state-of-the-art 
numerical calculations prove very difficult in regimes where the 
average number of fermions per site is different from 1. 
Unveiling the properties of the model in the latter regime, and 
addressing whether it supports superconductivity, may be crucial 
in understanding high-temperature superconductivity~\cite{e_dagotto_94}.

More than a decade ago, it was proposed that one could ``solve''
strongly-interacting quantum lattice models by ``simulating'' them 
using ultracold atoms in optical lattices~\cite{d_jaksch_98}. 
More recently, the Mott insulator in the Bose-\cite{m_greiner_02} 
and the Fermi-Hubbard~\cite{r_jordens_08} models were 
realized in these experiments. However, current accessible temperatures 
for fermions are still higher than one needs to observe even the 
relatively high temperature antiferromagnetic Ne\'{e}l transition in 
three dimensions.

These experiments are done using atoms with internal degrees of 
freedom, which can be selected to emulate not only two-specie fermions 
(Fermi-Hubbard model) and single-specie bosons (Bose-Hubbard model), 
but also particles with exotic statistics and/or pseudo-spins
\cite{i_bloch_08,g_thalhammer_08,s_trotzky_08,b_gadway_10,j_simon_11}. 
For instance, it has been shown that experiments with two-component 
(spin-1/2) bosons can lead to the realization of quantum spin models 
with tunable parameters~\cite{e_altman_03,l_duan_03,a_kuklov_03,
a_isacsson_05,a_hubener_09}. In the specific case where the intra-specie 
onsite repulsion is infinite, multiple occupancy of a single specie per 
site is forbidden. This case realizes an effective two-component hard-core 
boson (2HCB) model, which could be thought of as the bosonic equivalent of 
the Fermi-Hubbard model~\cite{e_altman_03,s_soyler_09,b_capogrosso_10}.
Similar to fermions, in the strong-coupling regime (large inter-species 
interactions), the low-energy properties of this model can be described 
by a $t$-$J$ model~\cite{m_boninsegni_01,y_nakano_11}.

Despite the outward similarities between the Fermi-Hubbard model and 
the 2HCB-Hubbard model (in one dimension they share identical thermodynamic
properties) in two dimensions, particle statistics plays a fundamental role 
in the properties of the system as the temperature is lowered, and in 
the selection of the ground state. At half filling, the ground state of 
fermions in two spatial dimensions is a Mott insulator with a
long-range Ne\'{e}l order for any value of the interaction strength, $U$. 
For 2HCBs, there is a quantum critical point at interaction 
$U_c/t\sim11$, where $t$ is the hopping amplitude, which 
separates a phase with two miscible strongly-interacting superfluids 
(2SF) at small inter-specie interactions from a Mott insulator 
super-counter-fluid (SCF)~\cite{a_kuklov_03} phase in the 
strong-coupling regime~\cite{s_soyler_09}. The latter state 
corresponds to a superfluid of pairs of bosons from one specie and 
holes of the other specie, and can also be interpreted as a 
long-range $XY$-ferromagnet in the iso-spin language. It is  
expected that such big contrasts in the nature of the ground states 
result in significant differences in the thermodynamic properties as 
well. Hence, for validating experiments with ultracold gases, which are 
performed at finite temperature, it is important to have 
access to exact quantitative results for the finite-temperature 
properties of the two systems.

While there have been numerous finite-temperature numerical studies of 
fermions in optical lattices~\cite{t_paiva_10,E_khatami_11b}, the same is
not true for 2HCBs, in particular in two dimensions. In one dimension, 
calculations were done introducing a generalized Jordan-Wigner 
transformation~\cite{s_chen_09}. The three-dimensional model with {\em attractive} 
interaction was studied to describe the supersolid state of 
$^4$He~\cite{x_dai_05,s_guertler_08}. Recently, dynamical mean-field 
theory results have been reported for the two-component (soft-core) 
Bose-Hubbard model in two dimensions with an average of 1/2 
particle of each specie per site and large intra-specie interactions 
\cite{a_hubener_09,y_li_12}. It was found in those studies that, upon heating 
from zero temperature, the system quickly enters an unordered Mott 
insulator. These findings have been complemented by finite-temperature 
quantum Monte Carlo (QMC) simulations of magnetic 
phases~\cite{b_capogrosso_10,t_ohgoe_11} as well as a field-theoretical
treatment~\cite{s_powell_09} in the hard-core limit. 

Here, we utilize numerical linked cluster expansions (NLCEs) to provide a 
comparative analysis between finite-temperature properties, such as the 
equation of state, entropy, specific heat, double occupancy, and spin 
correlations, of fermions and 2HCBs. We are particularly interested in 
identifying at what temperatures particle statistics become important 
in different interaction regimes, as well as what kind of qualitatively 
different behavior is produced by the statistics of the particles below 
those temperatures. We also discuss the implications of having bosonic vs
fermionic statistics for adiabatic cooling protocols and for detecting 
short-range spin correlations in ultracold atoms experiments. We note 
that the lowest temperatures that are accessible with NLCEs are typically 
higher than the crossover temperatures to the $XY$-ferromagnet 
phase~\cite{a_hubener_09}. Therefore, the behavior of the 
strongly-correlated systems that we study at half filling
is that of a Mott insulator with large spin 
correlations for either particle statistics.

We show that, in the weak-coupling regime ($U<U_c$), where the 
bosonic system has a superfluid ground state, the disagreement between 
properties of fermions and 2HCBs is apparent at relatively high 
temperatures (of the order of the hopping). In contrast, by increasing 
$U$ in the strong-coupling regime, the agreement between the thermodynamic 
properties of the two systems extends to lower temperatures. For 2HCBs, 
although the $z$-antiferromagnetic Ne\'{e}l ground state is known to win 
over the $XY$-ferromagnet only in cases where the two species have different 
hopping amplitudes, the two phases are very close in energy for equal 
hopping amplitudes (the case considered here) as only terms 
beyond the second order perturbation in interaction determine the 
difference~\cite{e_altman_03,a_kuklov_03}. Consistent with this picture, 
we find that at strong interactions, short-range $z$-antiferromagnetic 
correlations are large in the low-temperature Mott region and very close 
for the two particle statistics. 
Given that there are more efficient cooling techniques 
for bosons than for fermions, one could envision probing finite-temperature 
fermionic correlations using strongly correlated bosonic systems. Here, 
we present evidence that, through an adiabatic cooling mechanism 
that takes place in the bosonic system, the region with exponentially large 
antiferromagnetic (AF) correlations in two dimensions is more easily 
accessible with 2HCBs than with fermions.

The exposition is organized as follows: In Sec.~\ref{sec:model}, 
we introduce the model and discuss the NLCEs used in this study. We 
present the results for the 
thermodynamic properties of the models as well as their implications for 
the optical lattice experiment in Sec.~\ref{sec:res}. Our findings are
summarized in Sec.~\ref{sec:con}.

\section{the model and NLCEs}
\label{sec:model}

We consider the two-dimensional (2D) Hubbard Hamiltonian on the square 
lattice:
\begin{eqnarray}
\hat{H}&=&-t\sum_{\left <i,j\right >\sigma}(\hat{a}^{\dagger}_{i\sigma}
\hat{a}_{j\sigma} + \text{H.c.})+
U\sum_i \hat{n}_{i\uparrow} \hat{n}_{i\downarrow}\nonumber\\&&-\mu\sum_i 
(\hat{n}_{i\uparrow}+ \hat{n}_{i\downarrow})
\label{eq:ham}
\end{eqnarray}
where $\hat{a}^{\dagger}_{i\sigma}$ ($\hat{a}_{i\sigma}$) creates
(annihilates) a particle with spin $\sigma$ (for simplicity, we use 
spin instead of isospin for bosons too) on site $i$, and
$\hat{n}_{i\sigma}=\hat{a}^{\dagger}_{i\sigma} \hat{a}_{i\sigma}$ is the
number operator. $\langle..\rangle$ denotes nearest neighbors (NN),
and $U$ ($>0$) is the strength of the onsite repulsion. $t=1$ ($\hbar=1$ 
and $k_B=1$) sets the unit of energy throughout this paper. We consider
two different particle statistics: fermions
($\hat{a}_{i\sigma}=\hat{f}_{i\sigma}$) and 2HCBs
($\hat{a}_{i\sigma}=\hat{b}_{i\sigma}$). The 2HCB
operators satisfy the following commutation relations and constraints:
\begin{equation}
[\hat{b}^{}_{i\sigma},\hat{b}^{\dagger}_{j\sigma'}]=\delta_{ij}\delta_{\sigma\sigma'},
\qquad \hat{b}^{\dagger2}_{i\sigma}=\hat{b}^{2}_{i\sigma}=0.
\end{equation}

\subsection*{Numerical linked-cluster expansions}

We solve the Hamiltonian (\ref{eq:ham}) using the NLCEs introduced in 
Ref.~\cite{M_rigol_06}. In NLCEs, an extended property of the lattice model 
per site in the thermodynamic limit, $P$, is expanded in terms of contributions 
from all of the clusters, up to a certain size, that can be embedded in 
the lattice:
\begin{equation}
P=\sum_c L(c)w_p(c),
\label{eq:1}
\end{equation}
where $c$ represents the clusters. The contribution of each cluster with 
a particular topology is proportional to the number of ways it can be 
embedded in the lattice per site, $L(c)$, and its weight for the property of 
interest, $w_p(c)$. The weights are computed based on the 
inclusion-exclusion principle and given the property for each cluster, 
$\mathcal{P}(c)$, which is calculated using exact 
diagonalization~\cite{M_rigol_06}:
\begin{equation}
\label{eq:2}
w_p(c)=\mathcal{P}(c)-\sum_{s\subset c}w_p(s).
\end{equation}
Here, we
carry out the calculations up to the ninth order in the site expansion 
(maximum cluster size of nine sites).

NLCEs do not suffer from statistical or systematic errors, such as 
finite size effects and are not restricted to small or intermediate
interaction strengths. For this reason, they are complementary to
more commonly used methods, such as QMC simulations and 
dynamical mean-field theory, especially in the strong-coupling regime
($U$$\gg$$t$) where computations in the latter approaches become more
challenging. However, NLCE results are useful only in the temperature 
region in which the series converge, which has been shown to extend 
beyond the region accessible within high-temperature expansions 
\cite{M_rigol_06,grandcanonical}. The convergence of NLCEs can be 
accelerated by means of numerical resummations algorithms~\cite{M_rigol_06}. 
Here, we use Euler and Wynn methods with different parameters and plot the 
resulting last two orders (or only the last order when the two are indistinguishable 
in the figures). The results from all of those algorithms agree with each other 
in the regions shown within the small fluctuations seen in some cases at the 
lowest temperatures. These fluctuations, which occur below the convergence 
temperature of NLCE direct sums, arise from numerical 
instabilities in the resummations routines.

We work in the grand canonical ensemble \cite{grandcanonical}, and so, the 
exact diagonalization for every finite cluster in the series is performed 
in all particle and spin sectors. For each $U$, the partition 
function and all other observables are calculated in a dense grid of 
chemical potentials ($\mu$) and temperatures ($T$). This allows us to 
study their behavior at constant density $n=\langle
\hat{n}_{\uparrow}+\hat{n}_{\downarrow}\rangle=2\langle\hat{n}_{\sigma}
\rangle$ \cite{grandcanonical}, where $\langle .. \rangle$ denotes the 
expectation value. Since only NN hopping is considered, properties of 
the particle-doped system can be expressed in terms of those for the hole-doped 
system. Hence, in most cases away from half filling, we show results 
only for the hole-doped system ($n<1$).

\section{results}
\label{sec:res}

\subsection*{Entropy and specific heat}

Generally, when using QMC-based methods, the specific heat ($C_v$) or
the entropy ($S$) calculations involve numerical derivatives and/or
integration by parts~\cite{f_werner_05,k_mikelsons_09}, which can
introduce systematic errors. Within NLCEs, these two quantities are
computed directly from their definitions:
\begin{equation}
S=\ln(Z)+\frac{\langle
\hat{H}\rangle -\mu \langle \hat{n}\rangle }{T},
\end{equation}
where $Z$ is the partition function, and
\begin{equation}
C_v=\left(\frac{\partial \hat{\langle H\rangle}}{\partial
T}\right)_n=\left(\frac{\partial
\hat{\langle H\rangle}}{\partial T}\right)_{\mu}+\left(\frac{\partial
\hat{\langle H\rangle}}{\partial
\mu}\right)_T\left(\frac{\partial \mu}{\partial T}\right)_n.
\end{equation}
Since we work in the grand canonical ensemble, where the chemical 
potential and not the density is the control parameter, we have written 
this expression in a more suitable form for numerical evaluation. After
straightforward mathematical derivations, using Maxwell
equations, one can obtain the following closed form for the specific
heat in terms of expectation values that can be computed directly 
in NLCEs:
\begin{equation}
C_v=\frac{1}{T^2}\left[\langle\Delta
\hat{H}^2\rangle-\frac{
\left (\langle\hat{H}\hat{n}\rangle-\langle\hat{H}\rangle\langle\hat{n}
\rangle
\right)^2}{
\langle\Delta \hat{n}^2\rangle}\right],
\end{equation}
where $\langle \Delta \hat{H}^2\rangle = \langle \hat{H}^2\rangle
-\langle \hat{H}\rangle ^2$, and similarly $\langle \Delta
\hat{n}^2\rangle =\langle \hat{n}^2\rangle -\langle \hat{n}\rangle ^2$.

\begin{figure}
\centerline {\includegraphics*[width=3.3in]{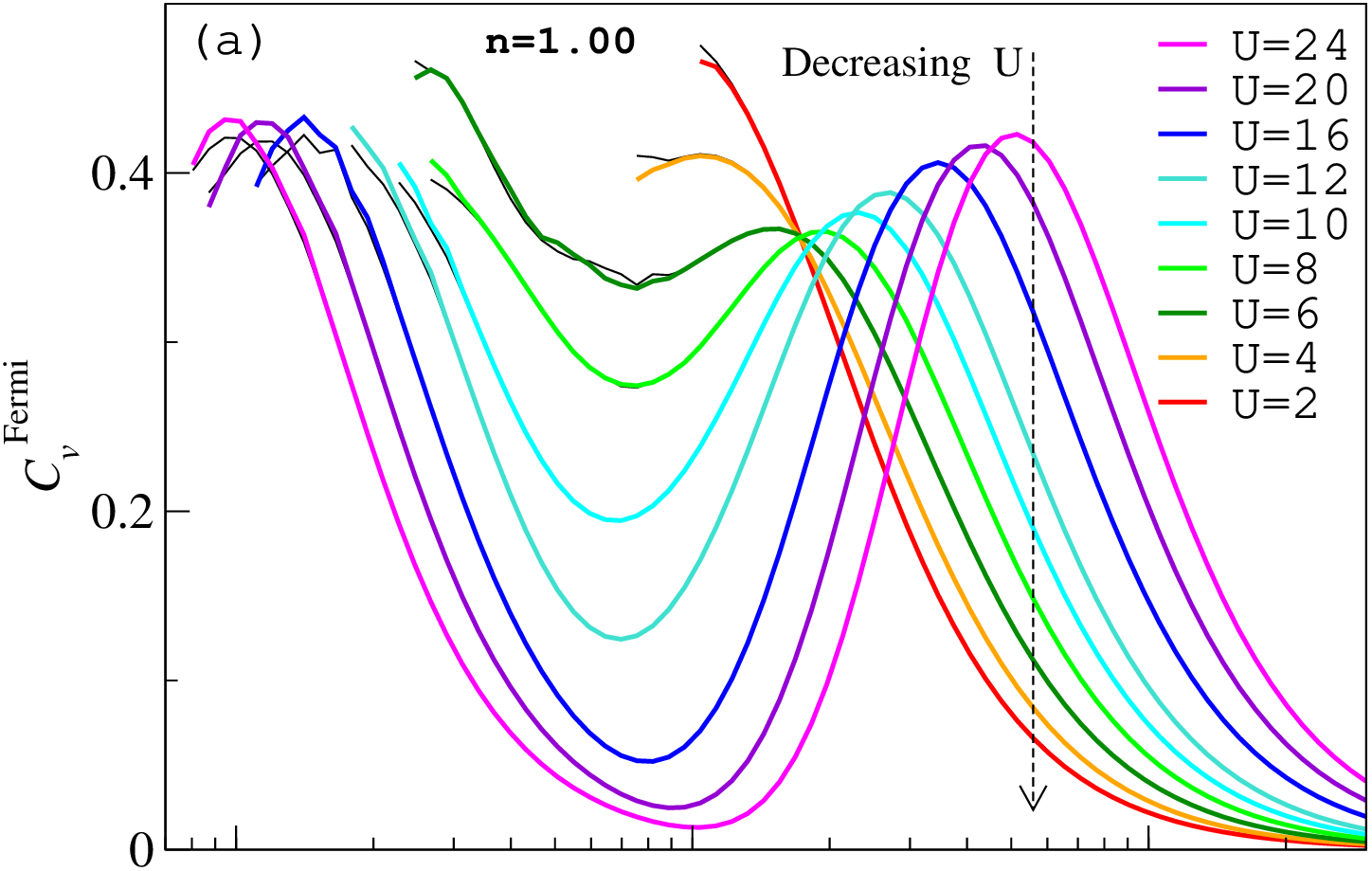}} 
\centerline {\includegraphics*[width=3.3in]{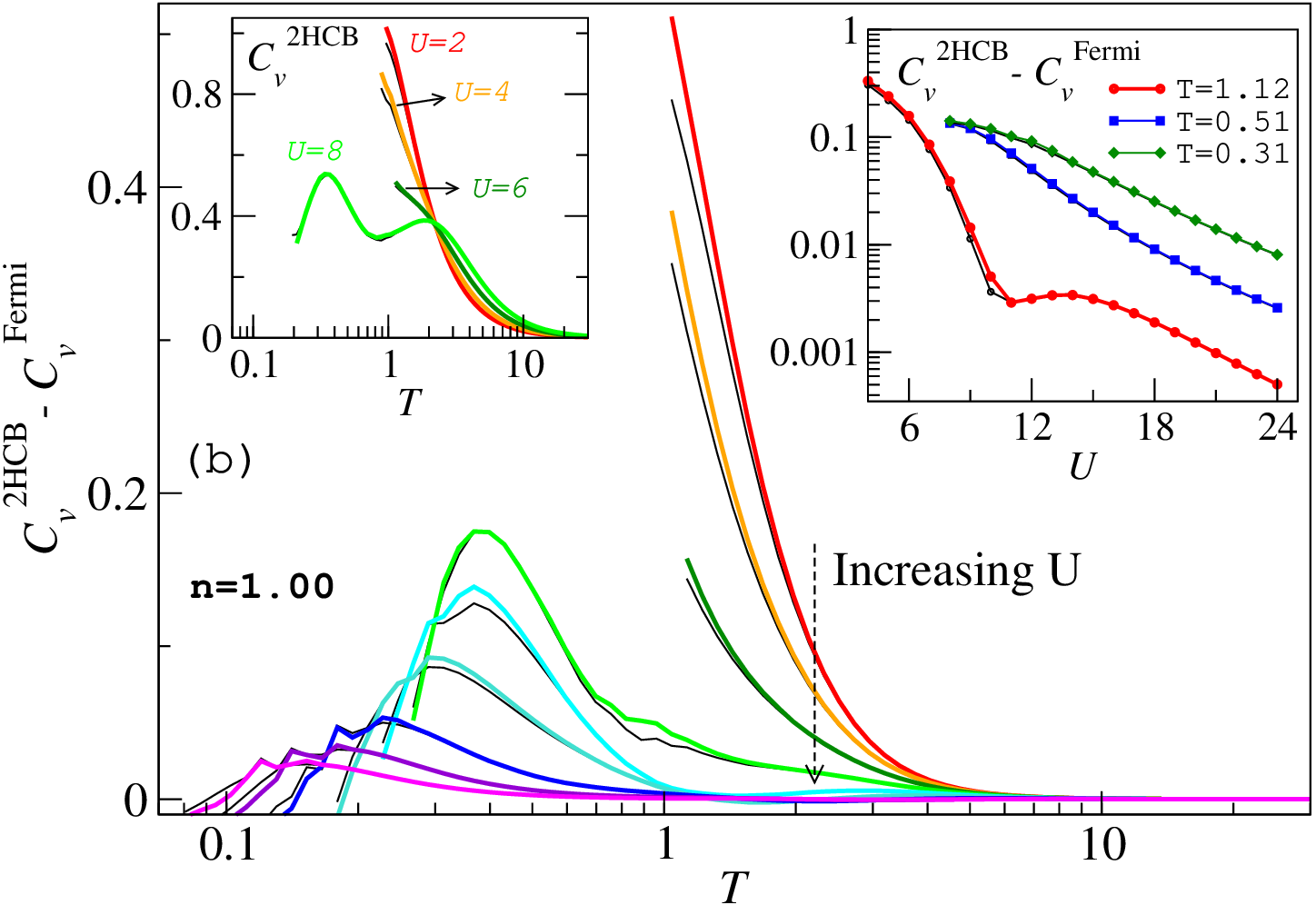}}
\caption{(Color online) (a) Specific heat ($C_v$) of the Fermi-Hubbard model
at half filling vs temperature for values of the onsite interaction
ranging from $U=2$ to three times the bandwidth ($U=24$). 
(b) Difference between the specific heat of the Fermi-Hubbard model 
and the 2HCB-Hubbard model at half filling vs temperature, for the same 
interactions as in (a). The left inset in (b) shows $C_v$ for 
the latter model vs $T$ for $U=2, 4, 6$ and $8$; and the right inset 
shows the difference between $C_v$ of the Fermi-Hubbard model 
and 2HCB-Hubbard model at half filling vs $U$ at three different 
temperatures. We have used 
Euler sums for the last $6$ terms. Thick (color) lines are the 
results of the sums up to the $9$th order and thin (black) 
lines up to the $8$th order.}
\label{fig:SH}
\end{figure}

We begin our study of the dependence of the observables on particle statistics 
as the temperature in the system is changed by showing, in Fig.~\ref{fig:SH},
the specific heat for fermions and 2HCBs. Figure \ref{fig:SH}(a) shows
the specific heat of the Fermi-Hubbard model from the weak-coupling 
regime to the strong-coupling regime with interactions up to three times
the bandwidth ($U=24$). The trend in the deviation of the specific heat
for 2HCBs from that of fermions for different interaction strengths is
depicted in Fig.~\ref{fig:SH}(b), where we show the difference between
the two. It is clear that for $U\lesssim 8$, the difference is
significant at temperatures greater than 1. For example, as shown in 
the left inset in Fig.~\ref{fig:SH}(b), the specific heat of 2HCBs for 
$U=2$ is roughly twice as large as that of fermions around $T=1$. Although
the exact trend of the former at lower temperatures cannot be resolved
within our method, such large high-temperature values in comparison to
those for the fermionic case are suggestive of the lack of a second
peak at lower temperatures in the 2HCB model. This is also supported
by a fast drop in the entropy (not shown). We have found that for $U=2$,
at $T\sim 0.8$, already $\sim 70\%$ of the infinite-temperature 
entropy has been quenched. For 2HCBs, a double-peak structure 
in the specific heat becomes apparent for $U\gtrsim 7$. At the same time, 
the minimum temperature at which NLCEs converge, which for smaller $U$ is 
generally higher for 2HCBs in comparison to fermions, extends to roughly 
the location of the low-temperature peak.

\begin{figure}[!t]
\centerline {\includegraphics*[width=3.3in]{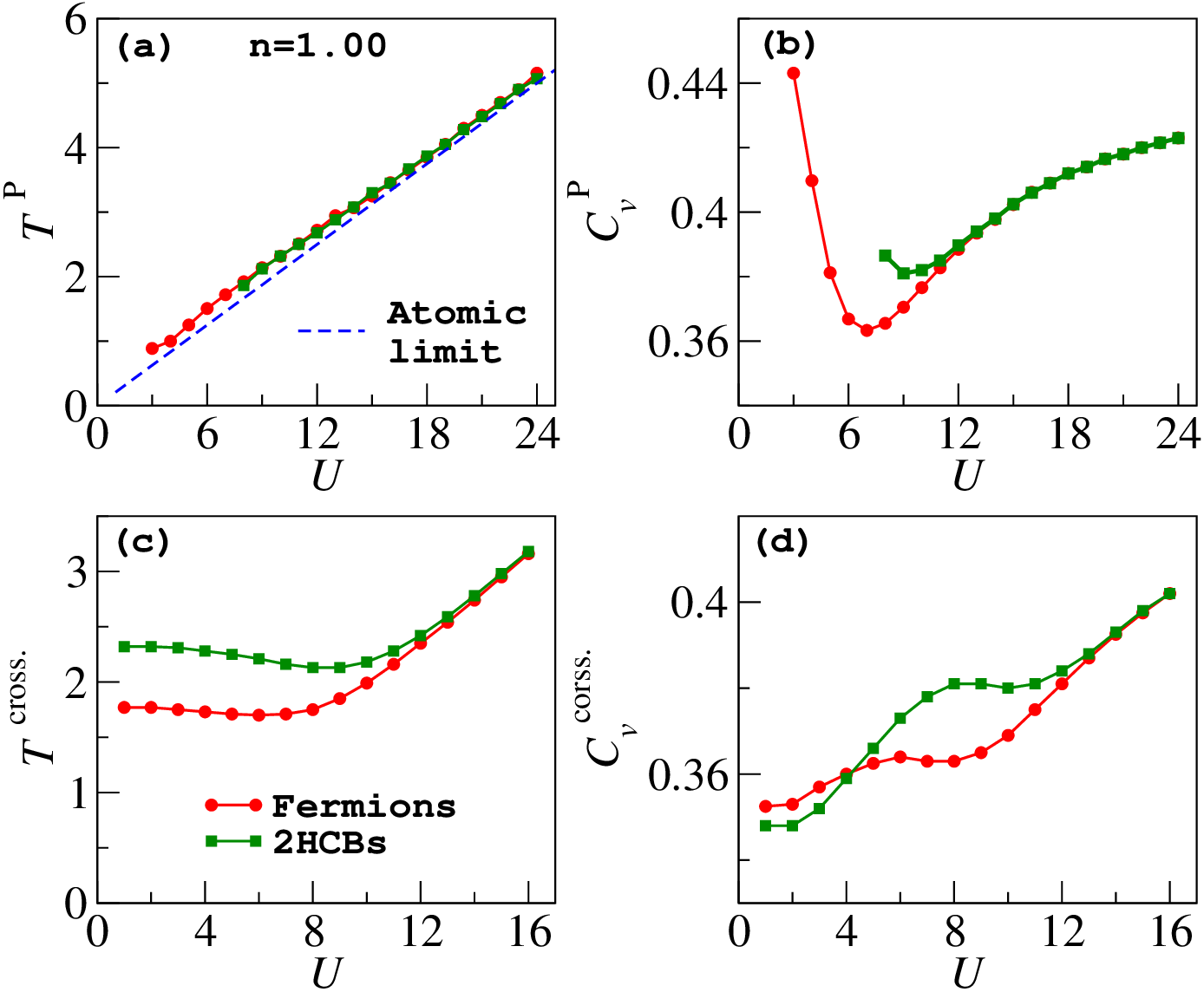}} 
\caption{(Color online) (a) Temperature and (b) value of the 
high-temperature peak of the specific heat ($C_v^{P}$) as functions of 
$U$. By decreasing $U$ in the strong-coupling regime, $C_v^{P}$ for the 
bosonic case deviates from the fermionic one
around $U=12$. For $U<8$, the two peaks in the specific heat of 2HCBs 
merge, and the high-temperature peak is not well-defined
anymore. (c) The location of the high-temperature crossing
point between the specific heat curves for consecutive values of $U$.
(d) The value of $C_v$ at that crossing point.}
\label{fig:peak}
\end{figure}

For fermions, our NLCE calculations resolve the double-peak structure of 
the specific heat in the thermodynamic limit for $U\geq 6$. Since QMC 
simulations can access lower temperatures for $U\lesssim 6$, 
previous QMC studies of this model have established the existence 
of the low-$T$ peak for any finite value of the interaction 
strength~\cite{t_paiva_01}, which signifies the crossover to the phase
with exponentially large AF correlations. In the strong-coupling regime, 
the specific heat results for fermions and 2HCBs are very close to each other 
at high temperatures ($T>1$). In this regime, the difference between the
$C_v$ of the two particle statistics decreases systematically for all 
accessible temperatures as $U$ is increased. As shown in the right inset 
in Fig.~\ref{fig:SH}(b), the reduction of this difference is nearly 
exponential for large values of $U$.

The high-$T$ peak, which is associated with the freezing of
charge degrees of freedom and moment formation, moves to higher $T$ as
$U$ increases. In the atomic limit ($t\rightarrow 0$), the location of
the peak, $T^P$, is determined by $\alpha \tanh(\alpha)=1$, where
$\alpha=U/16T^P$, which can be approximated by $T^P=U/4.8$. For
large $U$, and regardless of the particle statistics, we find a
very good agreement with the atomic limit prediction for $T^P$ 
[see Fig.~\ref{fig:peak}(a)]. Note that there is no well-defined
high-$T$ peak in the specific heat of 2HCBs for $U<7$. 
Unlike its position, the value of this peak does not 
change monotonically with increasing $U$ and, as seen in
Fig.~\ref{fig:peak}(b), has a minimum around $U=7$ for fermions and
around $U=9$ for 2HCBs. However, the relative change in the studied 
range of interactions is only about $20\%$. For large values of $U$,
the area under the high-$T$ peak of $C_v/T$ also approaches $\ln(2)$,
consistent with results in the atomic limit.

The low-temperature peak, which is associated with 
ordering of the moments, is expected to move to lower temperatures by
increasing $U$. This is because, for large values of $U$($\gtrsim 12$),
the low-$T$ system is essentially described by the
antiferromagnetic Heisenberg model with a characteristic energy
scale of $J\propto t^2/U$~\cite{E_khatami_11b}. Therefore, the position
of the low-$T$ peak, which we do not report here, 
is expected to be inversely proportional to $U$.

An interesting feature discussed in the past for correlated fermionic 
systems is the near universal high-$T$ crossing of the specific heat 
curves for different values of the interaction
\cite{d_duffy_97,d_vollhardt_97,t_paiva_01}. Due to its accuracy, 
NLCE provides an ideal tool for examining the precise behavior
of the crossing point. We show its position, and the value of 
$C_v$ at the crossing, in Figs.~\ref{fig:peak}(c) and \ref{fig:peak}(d), 
respectively. We find that, for fermions, the temperature and the specific 
heat value of the crossing point between two
consecutive values of $U$ is nearly independent of $U$ for $U\le 8$
($T^\text{cross.}\sim 1.75$, and $C_v^{\text{cross.}}\sim 0.36$), with
changes of roughly $5\%$. These variations are slightly larger for
2HCBs in the same range of interactions.


\begin{figure}[!t]
\centerline {\includegraphics*[width=3.35in]{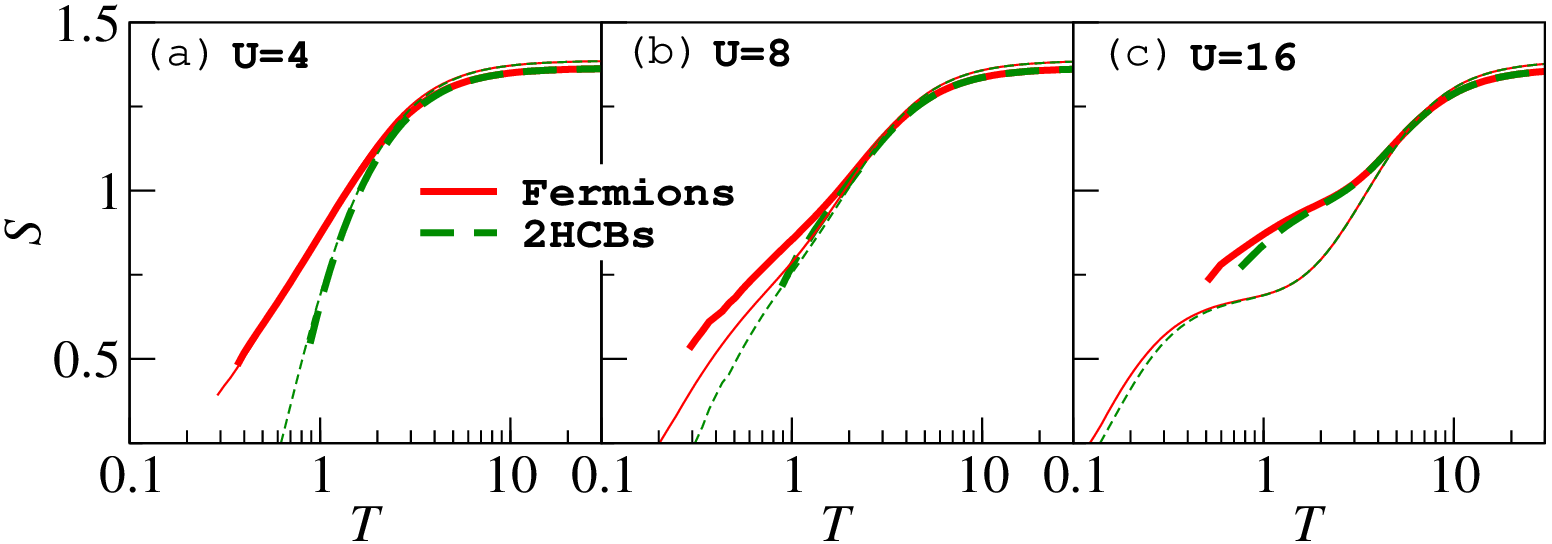}} 
\caption{(Color online) Comparison of the entropy vs temperature for 
$n=0.85$ (thick lines) and $n=1.00$ (thin lines) between the Fermi-Hubbard 
and the 2HCB-Hubbard models for (a) $U=4$, (b) $U=8$, and (c) $U=16$.}
\label{fig:EN}
\end{figure}

Away from half filling, where the ground state of neither system is known, 
the trends in the deviation of finite-temperature properties of 
fermions and 2HCBs is different from the one reported so far. As an example, 
we show in Fig.~\ref{fig:EN} the entropy (for $U=4$, 8, and 16) vs $T$ for 
$n=0.85$, and compare each curve with the one for $n=1.00$. Interestingly, 
in the weak-coupling regime, the entropy does not change 
significantly at the accessible temperatures for either particle 
statistics as one dopes the system away from half filling. For $U=8$ 
[Fig.~\ref{fig:EN}(b)], the entropy for the fermionic system with 15\% 
doping is larger than the half-filled value for $0.3<T<2$, while 
it does not change nearly as much with doping for 2HCBs for $T\gtrsim 0.9$.
By increasing the interaction to $U=16$, the agreement between the 
entropy of fermions and 2HCBs away from half filling improves, yet the 
deviations between the two for $n=0.85$ start at much higher temperatures in 
comparison to the half filled case [see Fig.~\ref{fig:EN}(c)]. These 
observations suggest that, even in the strong-coupling 
regime, the two systems away from half filling have fundamentally different  
phases.

\subsection*{Equation of state and compressibility}

\begin{figure}[!t]
\centerline {\includegraphics*[width=3.3in]{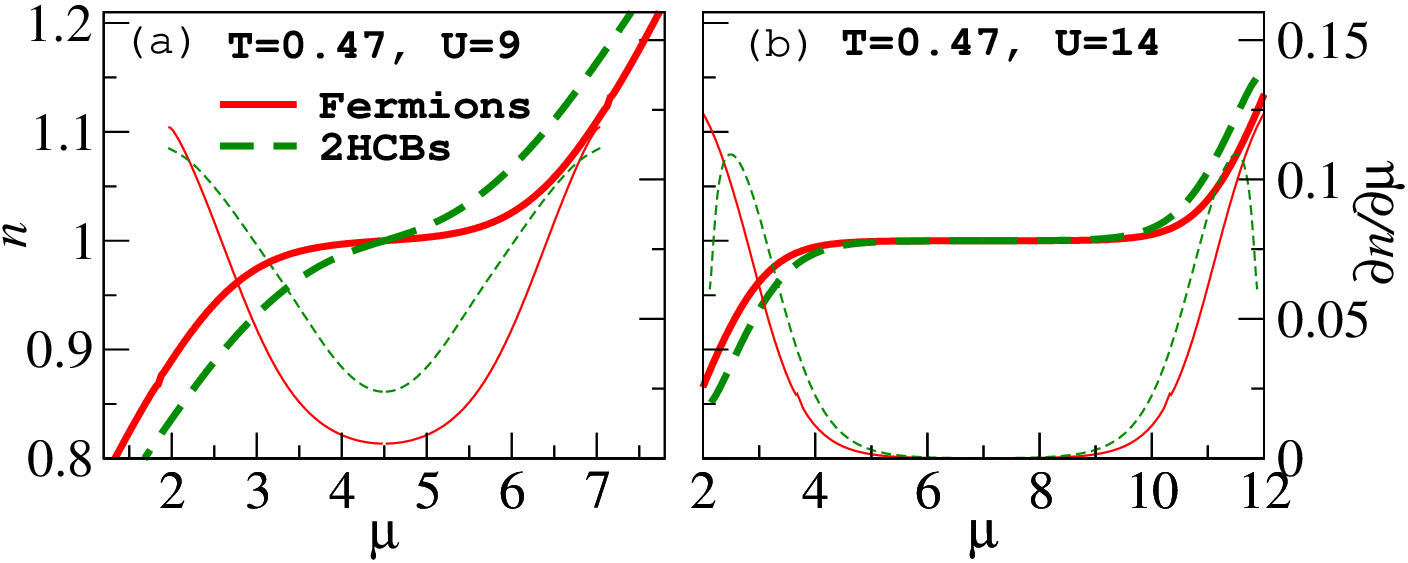}} 
\caption{(Color online) Equation of state and compressibility 
($\partial n/\partial \mu$) vs $\mu$ of fermions and 2HCBs for (a) 
$U=9$ and (b) $U=14$ at $T=0.47$. Thick lines show $n$ and thin 
lines show the compressibility. At zero temperature, the density of 
fermions is pinned at unity for $|\mu|<\Delta$, where $\Delta$ is the 
charge gap, regardless of $U$. For bosons, the system has no gap for 
$U< U_c$. Here, Wynn resummations with three cycles of improvement 
are used, and only the last order within the convergence region is shown.}
\label{fig:state}
\end{figure}

Further insight on the phases of the two systems at low temperatures can be 
gained by studying their equations of state and compressibilities. They are
also of great interest to optical lattice experiments since those
experiments are done in the presence of a confining potential that 
imposes a spatially varying chemical potential on the system. So, 
different regions in the trap correspond to different densities.
In Fig.~\ref{fig:state}, we show $n$ and $\partial n/\partial \mu$ 
vs $\mu$ at $T=0.47$ for $U=9$ and $U=14$, which are below and above 
the critical interaction value for 2HCBs. From QMC  
calculations it is known that, for $U<U_c$, the ground state of 2HCBs does 
not have a charge density gap~\cite{s_soyler_09}. One can also infer from 
Fig.~\ref{fig:state} that, regardless of the interaction strength,
2HCBs always have a smaller Mott gap at zero temperature than the fermions. 
This, in turn, implies that around half filling, the compressibility is 
always greater for bosons than for fermions, at any given temperature.

\subsection*{Double occupancy}

A slight upturn in the double occupancy ($D$) as one 
decreases $T$ for the half-filled strongly-correlated fermionic system 
is a known phenomenon that has been attributed 
to the increase in virtual hoppings between allowed nearest neighbor sites 
due to the enhancement of short-range AF correlations~\cite{e_gorelik_10}. In fact, the onset 
of this increase, which can be measured in the experiments, may serve 
as a universal probe for large AF correlations 
\cite{e_gorelik_11}. It can be shown to 
lead to adiabatic cooling by increasing the interaction strength~\cite{f_werner_05}, 
which is of great interest to optical lattice experiments. 
We have recently shown that such an increase in the 
double occupancy also occurs away from half-filling. Hence, in optical 
lattice experiments one also needs to make sure the density in most of the 
system is around half-filling in order for any increase in the double occupancy
to be associated with the onset of antiferromagnetism \cite{E_khatami_11b}.

\begin{figure}[!t]
\centerline{\includegraphics*[width=3.3in]{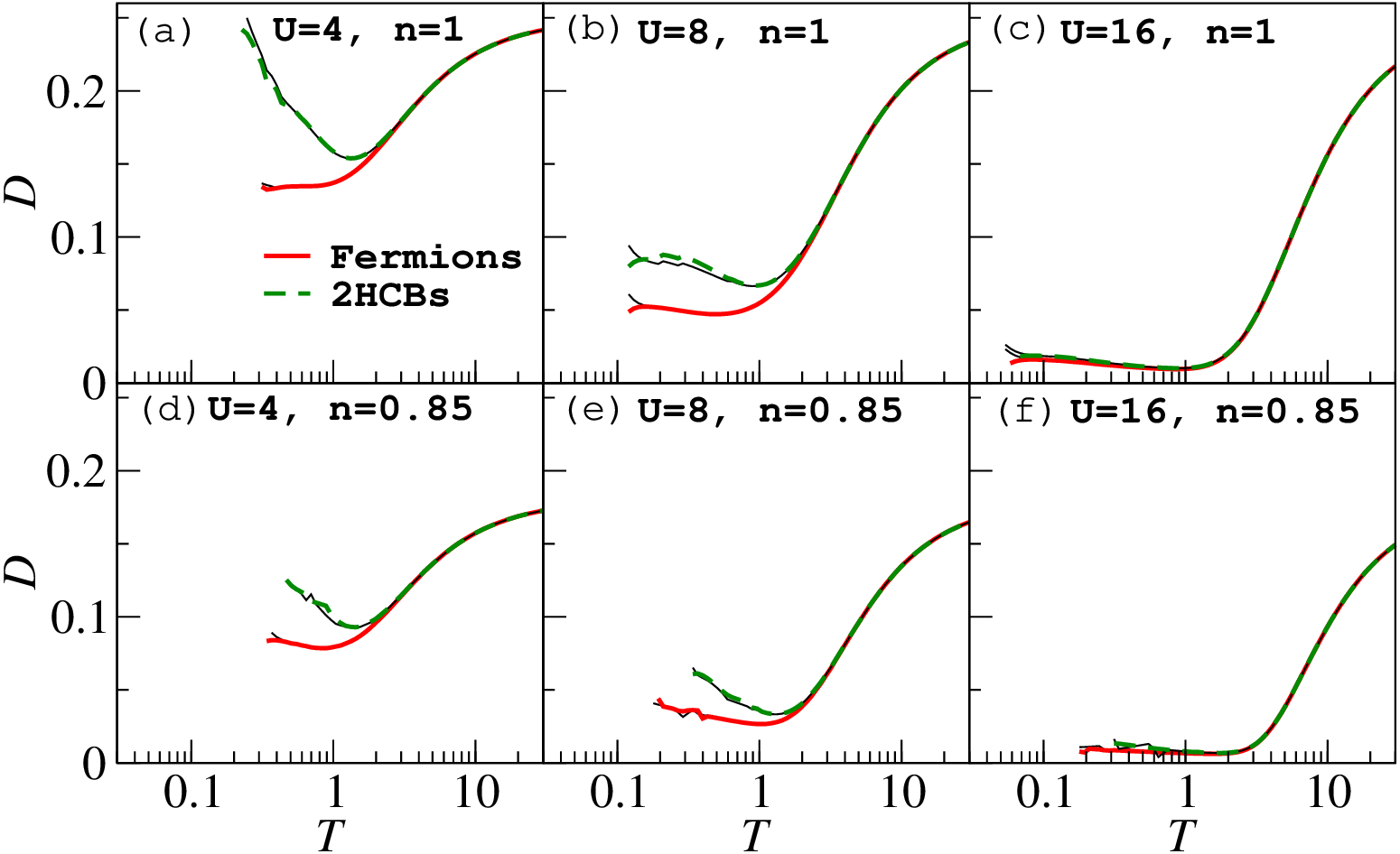}} 
\caption{(Color online) Double occupancy vs temperature for different 
values of $U$ at half filling (top panels) and for $n=0.85$ (bottom panels). 
In the weak-coupling regime with $U=4$ [(a) and (d)], $D$ for 2HCBs 
increases significantly by lowering the temperature below 1. Thick 
(color) lines are results from the last order and thin (black) lines are 
the results from the one-to-last order of the NLCEs after resummations.}
\label{fig:DO}
\end{figure}

We find interesting trends in the double occupancy of 2HCBs when compared 
to fermions, especially for weak interactions. 
As shown in Fig.~\ref{fig:DO}, and unlike in
the fermionic case, the double occupancy of 2HCBs at and away from half 
filling increases sharply below $T\sim1$ for weak interactions, e.g., $U=4$ 
in Figs.~\ref{fig:DO}(a) and \ref{fig:DO}(d), while such a large difference 
between the results for the two particle statistics is absent for large $U$. 
The sharp low-$T$ rise in $D$ indicates that adiabatic cooling starting from the 
weakly interacting limit is efficient for 2HCBs~\cite{f_werner_05}. 
The double occupancy for 2HCBs even 
reaches the uncorrelated value of $1/4$ in the half-filled case for $U=4$ and 
$T\sim 0.2$, in stark contrast to $D$ of the fermionic system. This is consistent 
with the absence of a Mott insulator for 2HCBs in that parameter region.

\subsection*{Spin correlations and uniform susceptibility}

\begin{figure}[!t]
\centerline{\includegraphics*[width=3.3in]{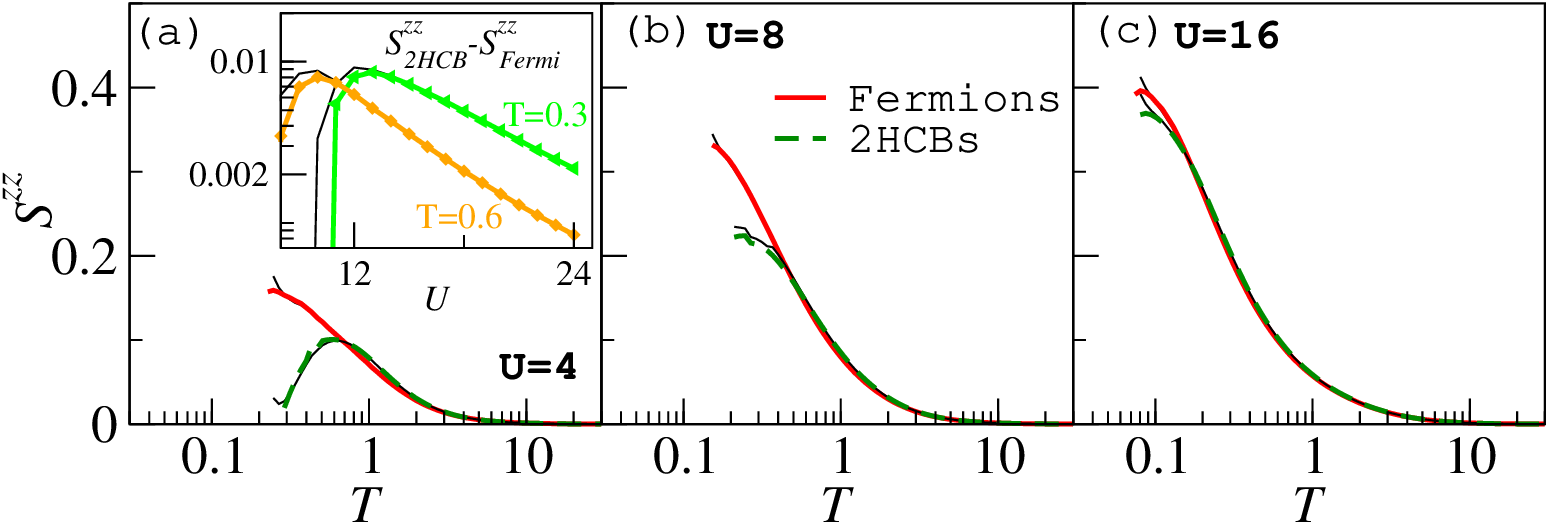}}
\centerline{\includegraphics*[width=3.3in]{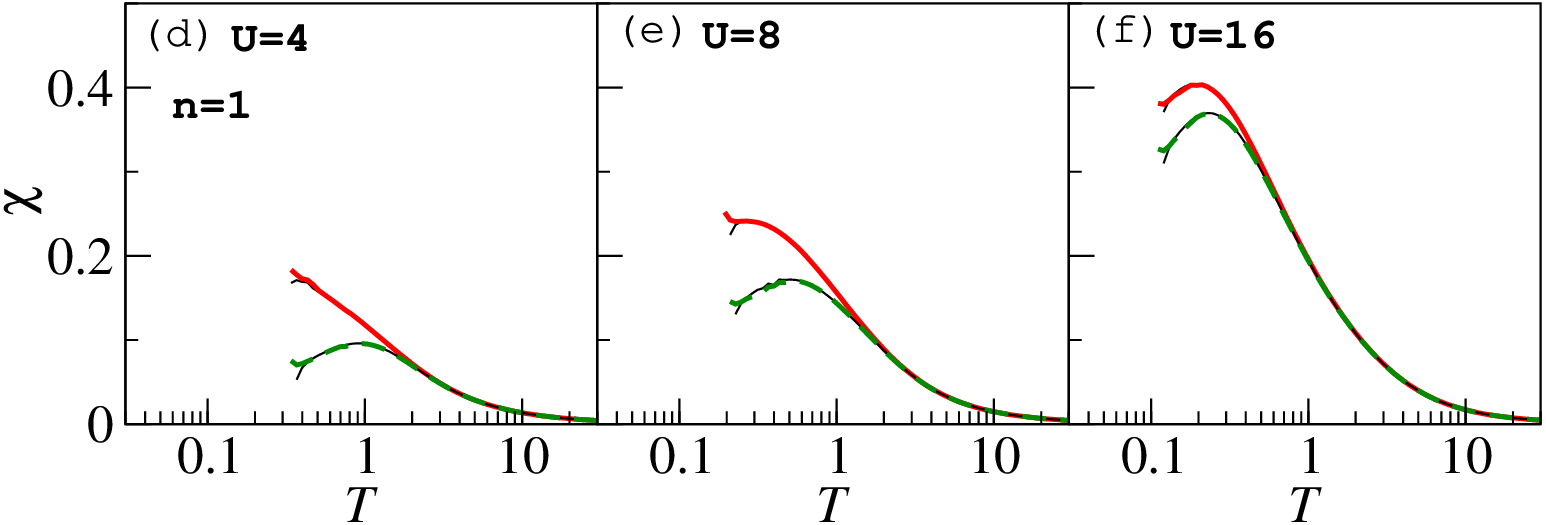}}
\caption{(Color online) Nearest-neighbor spin correlations vs temperature 
at half filling for (a) $U=4$, (b) $U=8$, and (c) $U=16$. (d)-(f)
Uniform spin susceptibility at half filling vs temperature. Like for
fermions, the spin susceptibility for 2HCBs peaks at a characteristic
temperature $T^*$. The inset in (a) shows the exponential decrease of 
the difference between $S^{zz}$ of 2HCBs and fermions at fixed temperatures
by increasing $U$ in the strong-coupling regime. Lines are the same as in
Fig.~\ref{fig:DO}}
\label{fig:SC}
\end{figure}

A related trend is also seen in the results for the NN spin correlations,
$S^{zz}=|\langle\sum_{\delta}S^z_i S^z_{i+\delta}\rangle|$, where the sum 
runs over the four nearest neighbors of site $i$. As depicted in Fig.~\ref{fig:SC}(a), 
$S^{zz}$ for 2HCBs peaks around $T=0.5$ when $U=4$ before becoming vanishingly small 
at lower temperatures, which is again consistent with the 
superfluid nature of the ground state in this interaction region.
On the other hand, it has been shown that for the fermionic case, $S^{zz}$ 
at half filling grows monotonically by decreasing the
temperature~\cite{t_paiva_01}, which is consistent with its 
antiferromagnetically ordered low-$T$ phase. It would be interesting to 
examine $S^{zz}$ at $T=0$ for 2HCBs and across the phase transition 
between the 2SF and the SCF phases, where $S^{zz}$ is presumably not small. 
Here, the difference between results for fermions and 2HCBs sets in 
around $T=1$ for $U=4$ and, like all other 
thermodynamic quantities at half filling, becomes smaller as $U$ is 
increased [see Figs.~\ref{fig:SC}(a)-(c)]. As shown in the inset of 
Figs.~\ref{fig:SC}(a), at fixed temperatures, this difference becomes exponentially 
small with increasing $U$ in the strong-coupling regime.

Another thermodynamic quantity that highlights the difference between 
fermions and 2HCBs in the weak-coupling regime is the uniform spin 
susceptibility ($\chi$). As seen in Fig.~\ref{fig:SC}(d)-(f), we 
find that the deviation of $\chi$ for 2HCBs from that of fermions is 
large at low temperature for $U=4$ and becomes smaller as $U$ increases 
to 8 and 16. More importantly, whereas a previous QMC study by Paiva {\it et al.}
\cite{t_paiva_10} has shown that the peak location in the fermionic case 
changes non-monotonically by increasing U (following the variations of the AF
correlations in the system), for 2HCBs, the peak temperature decreases
monotonically by increasing the interaction strength.
This is because, unlike for fermions, the peak in $\chi$ for 2HCBs in 
the weak-coupling regime does not signify moment ordering, but rather 
the disappearance of well-defined moments. This can be 
understood from the fact that NN spin correlations also decrease 
around the same temperature [see Fig.~\ref{fig:SC}(a)].

\subsection*{Attractive interactions}

It is known for the fermionic Hubbard model that there exists the 
following unitary particle-hole transformation~\cite{h_shiba_72} that 
takes the repulsive Hubbard model to the attractive one:
\begin{eqnarray}
\label{eq:trans}
\hat{f}_{j\uparrow} &=& \hat{d}_{j\uparrow}, \nonumber \\
\hat{f}_{j\downarrow} &=& e^{i(\pi,\pi)\cdot {\bf R}_j}
\hat{d}^{\dagger}_{j\downarrow},
\end{eqnarray}
where ${\bf R}_j$ is the displacement vector of site $j$. 
To see the effect of the transformation more clearly, it is easier to rewrite the 
Hamiltonian of Eq.~\eqref{eq:ham} in the so-called particle-hole
symmetric form:
\begin{eqnarray}
\hat{H}&=&-t\sum_{\left <i,j\right >\sigma}(\hat{f}^{\dagger}_{i\sigma}
\hat{f}_{j\sigma} + \text{H.c.}) \label{eq:ham2}\\&&+
U\sum_i \left(\hat{n}_{i\uparrow}-\frac{1}{2}\right)\left(\hat{n}_{i\downarrow}-
\frac{1}{2}\right) 
-\mu^{\prime}\sum_i (\hat{n}_{i\uparrow}+ \hat{n}_{i\downarrow}),\nonumber
\end{eqnarray}
where $\mu^{\prime}=\mu-\frac{U}{2}$. The transformation in Eq.~\eqref{eq:trans}
leaves the hopping term, as well as $n_{i\uparrow}$, invariant, but changes 
$n_{i\downarrow}$ to $1-n_{i\downarrow}$. As a result, the Hamiltonian in
Eq.~\eqref{eq:ham2} is transformed to:
\begin{eqnarray}
\hat{H}&=&-t\sum_{\left <i,j\right >\sigma}(\hat{d}^{\dagger}_{i\sigma}
\hat{d}_{j\sigma} + \text{H.c.})\\&&-
U\sum_i \left(\hat{n}'_{i\uparrow}-\frac{1}{2}\right)\left(\hat{n}'_{i\downarrow}-
\frac{1}{2}\right) 
-\mu^{\prime}\sum_i (\hat{n}'_{i\uparrow}- \hat{n}'_{i\downarrow})\nonumber
\label{eq:ham3}
\end{eqnarray} where $\hat{n}'_{i\sigma}=\hat{d}^{\dagger}_{i\sigma} 
\hat{d}_{i\sigma}$. At half filling ($\mu'=0$), the only change from
Eq.~\eqref{eq:ham2} will be the sign of the interaction $U$. 
Therefore, the energy spectral properties of the half-filled 
Fermi-Hubbard model, e.g., its specific heat, entropy, etc, are
invariant under $U\rightarrow -U$. The nature of the ground state,
however, is profoundly different in the repulsive and the attractive
models since this transformation maps charge correlations
to spin correlations and vice versa~\cite{j_hirsch_85}, which means
that long-range AF order is mapped to a charge-density-wave one.

A similar unitary transformation maps the repulsive Hubbard 
model for 2HCBs to an attractive one:
\begin{eqnarray}
\hat{b}_{j\uparrow} &=& \hat{d}_{j\uparrow} \nonumber \\
\hat{b}_{j\downarrow} &=& \hat{d}^{\dagger}_{j\downarrow}.
\end{eqnarray}
Therefore, similar to the fermionic case, the Hamiltonian is invariant 
under the change of sign of the interaction at half filling. 
The same argument presented above for the nature of the ground state of 
the fermionic model also applies to 2HCBs for $U>U_c$. In the regime 
$U<U_c$, we expect the 2SF phase to be the ground even with 
attractive interactions since, like the repulsive case, the system 
presumably gains more energy through the condensation of each specie 
than by minimizing the interaction energy.

\begin{figure}[!t]
\centerline{\includegraphics*[width=3in]{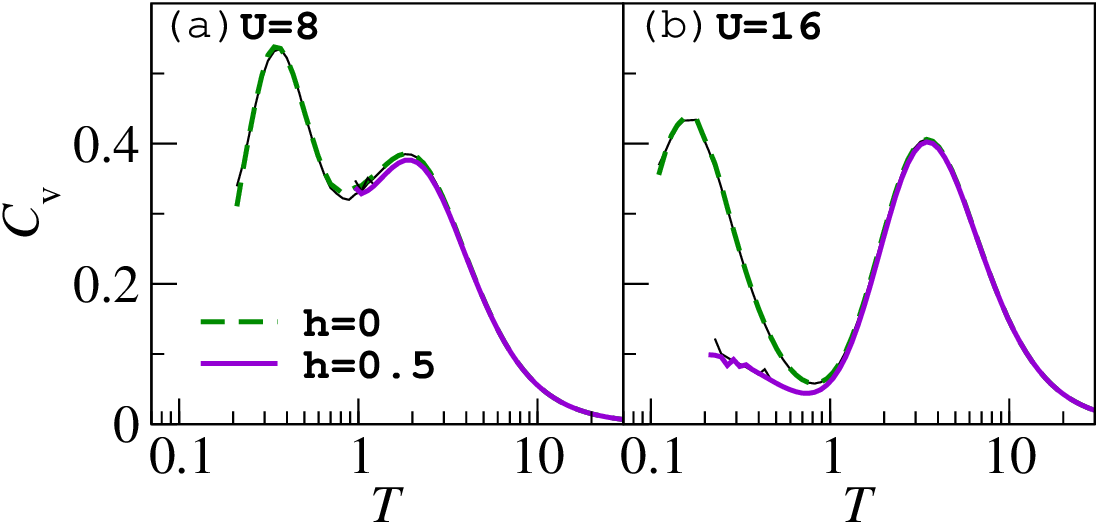}} 
\caption{(Color online) Specific heat of the repulsive Hubbard model 
for 2HCBs with and without a magnetic field, $h = 0.5$ and 0, for (a) $U=8$ and (b) 
$U=16$. Thick (thin) lines are the results from the last (one-to-last) 
order of the NLCEs after numerical resummation.}
\label{fig:attractive}
\end{figure}

As is clear from Eqs.~\eqref{eq:ham2} and \eqref{eq:ham3}, the attractive 
(repulsive) Hamiltonian away from half filling ($\mu'\neq 0$) is 
equivalent to the repulsive (attractive) one in the presence of a 
magnetic field in the $z$ direction, $h$, a role that is played by the 
chemical potential in the attractive (repulsive) case. 
In Fig.~\ref{fig:attractive}, we 
show how the specific heat of the repulsive Hubbard model for 2HCBs 
is modified in the presence of such a field, by plotting $C_v$ of the 
attractive model at $\mu'=0$ and $0.5$  ($h=0$ and 0.5 for the repulsive case). 
The low temperature region 
is not accessible to us at small $U$ and, in Fig.~\ref{fig:attractive}(a) 
for $U=8$, only a small deviation around the high-temperature peak can 
be seen when the magnetic field is introduced. 
The results for $U=16$ in Fig.~\ref{fig:attractive}(b) show a 
suppression of the specific heat at $T<1$, which are consistent with the 
fact that spin degrees of freedom emerge only in the latter temperature 
region. Results for the fermionic case show qualitatively the same
behavior as for 2HCBs for those two values of $U$.

\subsection*{Adiabatic cooling: \\implications for optical lattice experiments}

\begin{figure}[!t]
\centerline{\includegraphics*[width=3.33in]{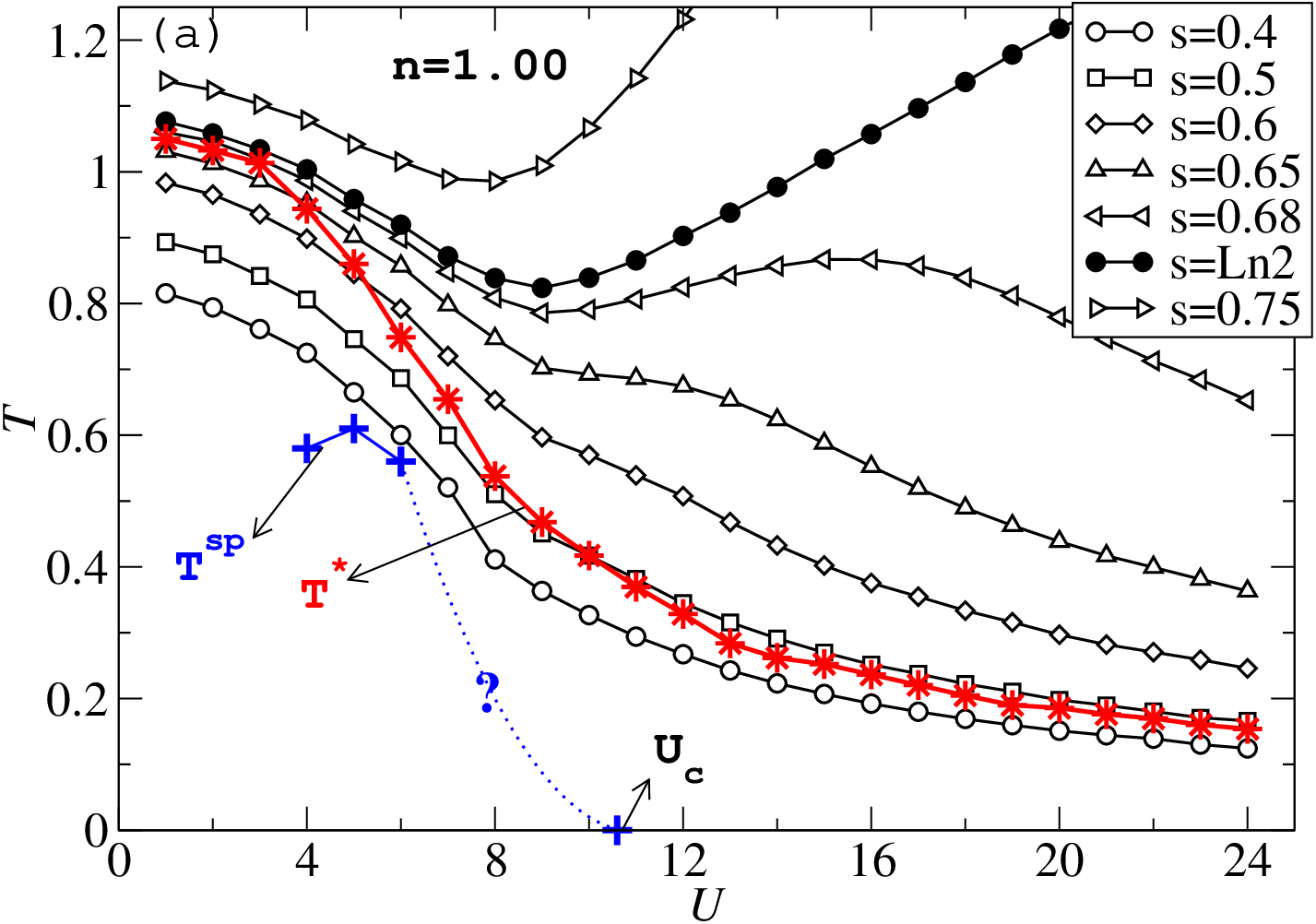}} 
\centerline{\includegraphics*[width=3.33in]{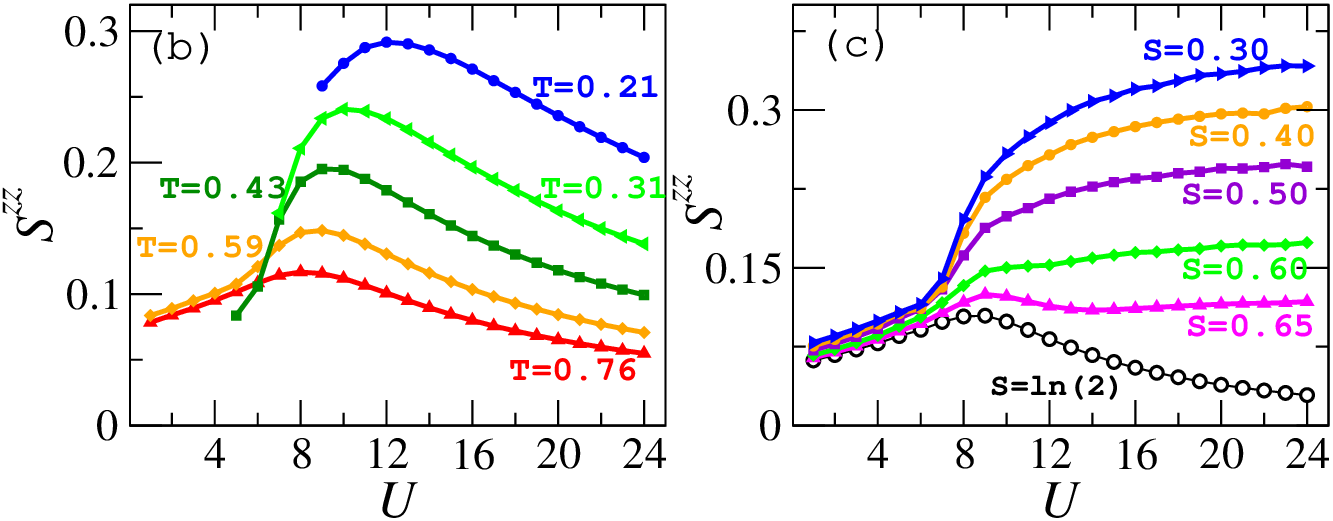}}
\caption{(Color online) (a) Isentropic dependence of temperature on the 
interaction for different values of the entropy for the 2HCB-Hubbard 
model at half filling. Also shown in this
panel is the characteristic temperature, $T^*$, which is the
location of the peak in the uniform spin susceptibility, and
$T^\text{sp}$, which is the location of the peak in the NN spin correlations.
The exact values of the latter are at lower temperatures than what is 
accessible to us for $U<4$ and $U>6$. (b) Isothermic curves of $S^{zz}$ 
vs $U$. (c) Isentropic curves of $S^{zz}$ vs $U$ for the entropy per 
particle from 0.3 to ln$(2)$.}
\label{fig:IST}
\end{figure}

Previously, we mapped out the isentropic paths of the 2D fermionic Hubbard model
in the extended temperature-interaction space~\cite{E_khatami_11b}. In 
Fig.~\ref{fig:IST}(a), we present a similar diagram for 2HCBs. As 
expected from the large negative slope in the low-$T$ double occupancy 
of 2HCBs for small $U$ vs $T$ (Fig.~\ref{fig:DO}), adiabatic cooling by 
increasing $U$ in the weak-coupling regime is much more 
efficient in comparison to the fermionic case. With the entropy per 
particle of 0.6, $T$ reduces roughly by a factor of 4 as the interaction
increases from 1 to 24. However, the underlying physics in different 
regions of the diagram and the change in the strength of AF 
correlations is unlike that of fermions. As mentioned before, in the 
fermionic system, regardless of the value of the interaction,  AF 
correlations are always enhanced by lowering the temperature. So, 
based on their dependence on $U$ and how rapidly the temperature falls 
in the adiabatic cooling process, one could drive the fermionic system 
into the region with exponentially large AF correlations. The latter 
region could be identified by the onset of a downturn in the uniform 
susceptibility ($T^*$)~\cite{E_khatami_11b}. 

On the contrary, for 2HCBs, the peak in the uniform susceptibility does 
not signify large AF correlations for weak interactions [see 
Figs.~\ref{fig:SC}(a) and \ref{fig:SC}(d)]. We emphasize this by 
plotting in the same figure the location of the peak in $S^{zz}$ ($T^\text{sp}$) 
for a few values of $U$. Beyond $U_c$, where the Mott insulating ground 
state not only has long-range $xy$-ferromagnetic order, but also very 
large $z$-antiferromagnetic correlations, the maximum of $S^{zz}$ is 
likely at zero temperature.

Results in Fig.~\ref{fig:IST}(b) and Fig.~\ref{fig:IST}(c) provide further 
insight into the behavior of $S^{zz}$ for different interaction strengths. By
comparing these figures to their fermionic counterparts [Figs. 3(c) and 3(d) 
in Ref.~\onlinecite{E_khatami_11b}], one can see that unlike in the weak-coupling 
regime, where the results are qualitatively different for the two particle statistics, 
in the strong-coupling regime they are very similar [as shown in Fig.~\ref{fig:SC}(a), 
the difference decreases exponentially with $U$]. As seen in Fig.~\ref{fig:IST}(b), 
at fixed (and low) temperatures 
$S^{zz}$ for 2HCBs sharply drops to small values by
decreasing $U$. In Fig.~\ref{fig:IST}(c), the 
isentropic curves of $S^{zz}$ are almost on top of 
each other for $U\lesssim 7$ for the entropies shown, and are expected to
become vanishingly small at lower entropies that are not accessible to us. 
This is, again, unlike the trends 
in the fermionic model in the weak-coupling regime. The opposite is true, 
however, for $U>U_c$ where, at the entropies accessible to us, $S^{zz}$ 
values of 2HCBs agree very well with their fermionic counterparts.

So far, much lower temperatures have been achieved with bosons in 
optical lattices than with fermions. Employing novel 
cooling techniques, researchers have been able to access temperatures as
low as 1 nK with bosons deep in the Mott insulating regime~\cite{d_weld_09}. 
It has also been shown in experiments with bosonic mixtures and that the 
inter- and intra-specie interactions can be tuned using Feshbach 
resonance and other techniques~\cite{g_thalhammer_08,b_gadway_10,j_simon_11}. 
Our results show that thermodynamic properties and short-range 
AF correlations of 2HCBs are very (exponentially) close to those of the 
fermions for strong interactions and up to intermediate to low temperatures. 
Therefore, by engineering the strong interaction regime studied here, optical 
lattice experiments with two-component bosons could also provide a tool for 
simulating intermediate to low temperature correlations in the Fermi-Hubbard 
model. Further studies need to be done to explore the properties of 2HCBs 
away from half filling and examine their relevance to the fermionic case.

\section{Conclusions}
\label{sec:con}

We have examined the particle statistics dependence of the thermodynamic 
properties of the 2D Hubbard model by means of an exact method (NLCEs) 
that yields properties such as entropy, specific heat, double occupancy,
and spin correlations in the thermodynamic limit. The results 
are valid above a certain temperature that depends on the interaction 
and the filling factor. We considered two-component fermions and 2HCBs 
and compared their properties at various temperatures and interaction 
strengths, up to three times the bandwidth. 

We have shown that for weak interactions, in the regime where the ground 
state of the two models at half filling have fundamentally different 
natures, the results for observables at finite temperature differ significantly 
for the two particle statistics starting at relatively high temperatures. 
In contrast, in the strong-coupling 
regime (beyond a critical interaction that separates a superfluid 
from a Mott insulator phase for 2HCBs), the agreement between the thermodynamic 
quantities of the two systems, including short-range AF correlations,
extends to much lower temperatures. We find that the trends in spin
correlations in the $z$ direction are similar for both particle statistics 
and that the results differ only by exponentially small values for strong 
interactions. This provides an additional tool to probe correlations of 
fermionic systems in optical lattice experiments by emulating them using 
two-specie bosons, which can generally be cooled down to lower temperatures.

\section*{Acknowledgments}

This work was supported by the NSF under Grant No.
OCI-0904597. We thank J. Carrasquilla for useful 
discussions.


\end{document}